\documentclass[12pt,preprint]{aastex}

\shorttitle{}

\shortauthors{Liu et al.}

\begin{document}

\title{Propagation of the 2012 March Coronal Mass Ejections from the Sun to Heliopause}

\author{Ying D. Liu\altaffilmark{1}, John D. Richardson\altaffilmark{2}, 
Chi Wang\altaffilmark{1}, and Janet G. Luhmann\altaffilmark{3}}

\altaffiltext{1}{State Key Laboratory of Space Weather, National
Space Science Center, Chinese Academy of Sciences, Beijing 100190, China; 
liuxying@spaceweather.ac.cn}

\altaffiltext{2}{Kavli Institute for Astrophysics and Space Research, 
Massachusetts Institute of Technology, Cambridge, MA 02139, USA}

\altaffiltext{3}{Space Sciences Laboratory, University of California, 
Berkeley, CA 94720, USA}

\begin{abstract}

In 2012 March the Sun exhibited extraordinary activities. In particular, the active region NOAA AR 11429 emitted a series of large coronal mass ejections (CMEs) which were imaged by STEREO as it rotated with the Sun from the east to west. These sustained eruptions are expected to generate a global shell of disturbed material sweeping through the heliosphere. A cluster of shocks and interplanetary CMEs (ICMEs) were observed near the Earth, and are propagated outward from 1 AU using an MHD model. The transient streams interact with each other, which erases memory of the source and results in a large merged interaction region (MIR) with a preceding shock. The MHD model predicts that the shock and MIR would reach 120 AU around 2013 April 22, which agrees well with the period of radio emissions and the time of a transient disturbance in galactic cosmic rays detected by Voyager 1. These results are important for understanding the ``fate" of CMEs in the outer heliosphere and provide confidence that the heliopause is located around 120 AU from the Sun.
 
\end{abstract}

\keywords{shock waves --- solar-terrestrial relations --- solar wind --- Sun: coronal mass ejections (CMEs) --- Sun: heliosphere}

\section{Introduction}

Coronal mass ejections (CMEs) are large-scale expulsions of plasma and magnetic field from the solar atmosphere, which can cause significant disturbances in the heliosphere. While the origin of CMEs and their propagation in the inner heliosphere have been extensively studied, evolution of CMEs in the outer heliosphere is not well understood. This is a particularly intriguing question concerning the ``fate" of CMEs in the heliosphere.  

A few large CMEs have been traced to the outer heliosphere using MHD propagation of observed solar wind disturbances \citep{wang01, wang02, richardson02, richardson05, richardson06, liu08, liu11}. A series of CMEs can produce a merged interaction region (MIR) in the outer heliosphere \citep{burlaga84, wang02, richardson02}, a shell of disturbed material with enhanced magnetic fields which can act as a barrier for cosmic ray transport \citep{burlaga85}. An MIR is often preceded by an interplanetary shock. Previous studies suggest that MIRs and their associated shocks resulting from stream interactions are a major mechanism modulating the outer heliosphere \citep{whang85, whang91, whang99, burlaga03a, burlaga03b}. When the shock and MIR impact the heliopause, the boundary between the heliosphere and interstellar space, they can generate heliospheric radio emissions observable by the Voyager spacecraft \citep[e.g.,][]{gurnett93, gurnett03}. 

The location of the heliopause is of particular interest. On 2012 August 25 at a distance of about 122 AU, Voyager 1 (V1) detected a dramatic decrease in anomalous cosmic rays while a simultaneous increase in galactic cosmic rays \citep{webber13, krimigis13}. This indicates that V1 has crossed a boundary, termed the ``heliocliff", which is thought to be related to the heliopause. Magnetic field observations, however, show that the magnetic field is essentially along the Parker spiral direction \citep{burlaga13a}. The absence of an appreciable change in the magnetic field direction has resulted in doubts regarding whether the ``heliocliff" is the heliopause. A new name was coined, the heliosheath depletion region, for the region that V1 has entered \citep{burlaga13a, stone13}. A definitive determination of the region's nature lies in measurements of the plasma density, which is believed to be much higher in interstellar space ($\sim$0.1 cm$^{-3}$) than in the heliosheath ($\sim$0.002 cm$^{-3}$). V1 does not have a working plasma instrument, but heliospheric radio emissions were observed by V1 from 2013 April 9 to May 22, which are hypothesized to have been produced by the 2012 March CMEs hitting the heliopause and interacting with the interstellar plasma \citep{gurnett13}. With this unexpected ``gift" from the Sun, the plasma density is deduced to be about 0.08 cm$^{-3}$, which suggests that the ``heliocliff" is the heliopause. 

In 2012 March the Sun indeed exhibited substantial activities including a series of M/X class flares and large CMEs \citep{liu13}. They were imaged by the wide-angle cameras aboard the Solar Terrestrial Relations Observatory \citep[STEREO;][]{kaiser08}, and caused significant solar wind disturbances observed at Wind. The purpose of this Letter is twofold: (1) to investigate how the CMEs propagate from the Sun to the outer heliosphere; and (2) to test the location of the heliopause. The results obtained here are crucial for understanding the ``fate" of CMEs in the outer heliosphere as well as the dimensions of the heliosphere. 

\section{Observations and Results}

Figure~1 shows the configuration of the planets and spacecraft in the ecliptic plane as well as the propagation directions of the major CMEs in 2012 March. STEREO A and B were about $109^{\circ}$ west and $118^{\circ}$ east of the Earth, respectively. Mars was about 1.66 AU from the Sun and a few degrees east of the Earth, while Saturn was about 9.7 AU from the Sun and $39^{\circ}$ west of the Earth. Voyager 2 (V2) and V1 were at distances of 98 and 120 AU from the Sun, heliographic latitudes of $-30^{\circ}$ and $34.5^{\circ}$, and longitudes of $126.5^{\circ}$ and $83^{\circ}$ west of the Earth, respectively. The CMEs shown in Figure~1 were all produced from the same active region, NOAA AR 11429, as it rotated with the Sun from the east to west: a CME of about 1500 km s$^{-1}$ associated with an X1.1 flare from N17$^{\circ}$E52$^{\circ}$ around 04:00 UT on March 5 (CME0); a CME of about 2500 km s$^{-1}$ associated with an X5.4 flare from N17$^{\circ}$E21$^{\circ}$ around 00:20 UT on March 7 (CME1); a CME of over 1500 km s$^{-1}$ associated with an X1.3 flare which occurred about 1 hour later than CME1; a CME of about 1000 km s$^{-1}$ associated with an M6.3 flare from N17$^{\circ}$W03$^{\circ}$ around 03:50 UT on March 9; a CME of about 1500 km s$^{-1}$ associated with an M8.4 flare from N17$^{\circ}$W24$^{\circ}$ around 17:40 UT on March 10 (CME2);  and a CME of about 1800 km s$^{-1}$ associated with an M7.9 flare from N17$^{\circ}$W61$^{\circ}$ around 17:30 UT on March 13 (CME3).

The turbulent corona and inner heliosphere can be seen from the time-elongation maps in Figure~2, which are produced by stacking the running-difference images from STEREO within a slit along the ecliptic plane \citep[e.g.,][]{sheeley08, davies09, liu10a, liu10b}. Several tens of CME occurred from 2012 March 1 to 20 (see the supplementary animation), but they do not necessarily leave tracks in the ecliptic time-elongation maps. However, the major eruptions, especially those that impacted the Earth, are all revealed by the maps. Comparing the tracks to the observed shock arrival times at the Earth, we can establish the connections of those CMEs (specifically, CME0, CME1, CME2 and CME3) with their near-Earth in situ signatures. We can also determine the CME kinematics using a triangulation technique based on the time-elongation maps \citep{liu10a, liu10b, lugaz10, davies13}. For instance, the Sun-to-Earth propagation of CME1 (a typical fast event) shows three phases: an impulsive acceleration, then a rapid deceleration, and finally a nearly constant speed propagation \citep{liu13}. This is not a focus of the present work though. Note that some of the tracks cross, which implies CME-CME interactions.

The in situ signatures at Wind are shown in Figure~3. Three interplanetary CMEs (ICMEs), the in situ counterparts of CMEs, are identified during the time period using the depressed proton temperature (complemented with rotation in the magnetic field). The weak shock at 03:22 UT on March 7 (S0) seems driven by CME0, whereas no driver signatures are observed. The CME propagation direction (E52$^{\circ}$) could be so far east that only the shock is observed at Wind (see Figure~1). ICME1 and its shock at 10:19 UT on March 8 (S1) are produced by CME1, ICME2 and its shock at 08:10 UT on March 12 (S2) by CME2, and ICME3 and its shock at 12:29 UT on March 15 (S3) by CME3, as is evident from Figure~2. The short interval of ICME3 is consistent with the CME propagation direction (W61$^{\circ}$; see Figure~1), i.e., only the flank is observed at Wind. The event following CME1 on March 7 and the CME from March 9 (03:50 UT), which could hit the Earth, do not have in situ signatures at Wind. Their signatures could have been lost during CME-CME interactions or interactions with the ambient solar wind. 

The clustering of ICMEs and shocks in Figure~3 indicates interactions between them, which could result in formation of a global MIR in the outer heliosphere. We propagate the solar wind disturbance outward from 1 AU using an MHD model \citep{wang00}, in an effort to look at the evolution in the outer heliosphere and make predictions that may be compared with measurements beyond 1 AU. The effects of pickup ions are included (e.g., solar wind heating and slowdown), so the model can propagate solar wind measurements to any distance in the heliosphere. The model has had success in connecting solar wind observations at different spacecraft \citep[e.g.,][]{wang01, wang02, richardson02, richardson05, richardson06, liu06, liu08}. A direction impression from Figure~1 is to use solar wind measurements at STEREO A as input to the model, as V1 is closer to STEREO A than the Earth. However, no clear ICME signatures are observed at STEREO A during the same period as in Figure~3, and the solar wind speed there is generally below 500 km s$^{-1}$. The Wind data may be more representative of the situation at V1's longitude than the STEREO A measurements, and are thus used as input to the model. Note that what would reach V1 is probably the flank of the MIR/shock resulting from the CME interactions.

Figure~4 shows the evolution of the solar wind streams from the Earth to Saturn. The three adjacent transient streams associated with the ICMEs interact with one another, during which the middle one first ``damps out" by momentum exchange with the surrounding flows and then a new one emerges. The predicted arrival time of the first stream at Mars is about 07:12 UT on March 8, while Saturn would encounter the first stream around 14:24 UT on March 30. Figure~5 puts the streams in a wider context. By the time they reach 20 AU, the three neighboring flows appear entrained in a bigger one. At 30 AU they have coalesced and formed a large shock followed by an MIR. The shock and MIR persist further out, although their energy decays with small flows being ``peeled off" from the stream. At 80 AU the MIR, within which the magnetic field is enhanced above the upstream ambient value, has a radial width of about 8 AU. As suggested by \citet{burlaga85}, the MIR could be a barrier for cosmic ray transport toward the Sun. Obviously, the identities of the original streams have been lost in the outer heliosphere, and the ``fate" of the CMEs is an MIR preceded by a shock which would finally hit the heliopause. 

Of particular interest is when the shock and MIR will reach the heliopause or the location of the heliopause. The model predicts that the shock arrives at 120 AU around 2013 April 22 (day 478 in Figure~5), which lies in the interval of the radio emissions detected by V1, i.e., 2013 April 9 - May 22 \citep{gurnett13}. The shock arrival time at 120 AU is also close to the period of a transient disturbance in galactic cosmic rays observed by V1, i.e., 2013 March 21 - April 5 \citep{krimigis13}. This supports the postulation of \citet{krimigis13} that the disturbance in galactic cosmic rays is associated with a large MIR generated by the 2012 March eruptions. If the interpretation that the radio emissions were produced by the shock and MIR hitting the heliopause is correct, then a direct inference from the good timing would be that the heliopause is around 120 AU from the Sun. This agrees very well with the heliopause location determined from the plasma density \citep{gurnett13} and the distance where cosmic rays show abrupt changes \citep{webber13, krimigis13}.

Note that the model we use does not include the transition across the termination shock and the heliosheath. The bulk of the solar wind energy in the heliosheath resides in the pickup ions, which are not measured, so we cannot determine the fast-mode speed in the heliosheath. Instead we look at propagation of shocks through the Earth's bow shock and magnetosheath as an analogy. Propagation of shocks through the Earth's magnetosheath has been studied using multiple spacecraft, at least one to identify shocks in the solar wind and one to observe them in the magnetosheath. Estimates for the slowdown of the shock are derived by the timing of the shock passage at each spacecraft, ranging from 0.7 - 1 of the upstream shock speed \citep[e.g.,][]{szabo03, koval06, pallocchia10}. If a shock in the heliosheath has similar decelerations as in the Earth's magnetosheath, we would expect a slowdown of 10 - 30\% from the termination shock (94 AU) to the heliopause. Figure~5 shows that the propagation speed in the model is $\sim$0.25 AU per day from 100 to 120 AU. Thus a 10 - 30\% slowdown in the heliosheath will increase the shock propagation time by 10 - 35 days, so the shock location would be consistent with the end of the radio emissions (as it was in the 2012 shock and radio emission case described by \citet{burlaga13b}).

\section{Summary}

We have investigated the propagation of the major CMEs of 2012 March through the heliosphere, using wide-angle imaging observations, in situ measurements and an MHD model. The CMEs produce a cluster of shocks and ICMEs at 1 AU. They interact with one another and finally in the outer heliosphere form a large MIR with a preceding shock which would eventually hit the heliopause. Memory of the source information about the individual events is lost during this long journey to the heliopause. As the shock and MIR sweep through the heliosphere, a significant portion of the heliospheric plasma will be shocked. The predicted arrival time of the shock and MIR at 120 AU is around 2013 April 22, consistent with the period of radio emissions and the time of a transient disturbance in galactic cosmic rays observed by V1 \citep{gurnett13, krimigis13}. This supports the view that V1 crossed the heliopause into interstellar space at a distance of about 120 AU from the Sun \citep{gurnett13}.

\acknowledgments The research was supported by the Recruitment Program of Global Experts of China, by the SPORT project under grant XDA04060801, by the Specialized Research Fund for State Key Laboratories of China, by the CAS/SAFEA International Partnership Program for Creative Research Teams, and by the STEREO project under grant NAS5-03131. We acknowledge the use of data from STEREO and Wind.

\clearpage

\begin{figure}
\epsscale{0.9} \plotone{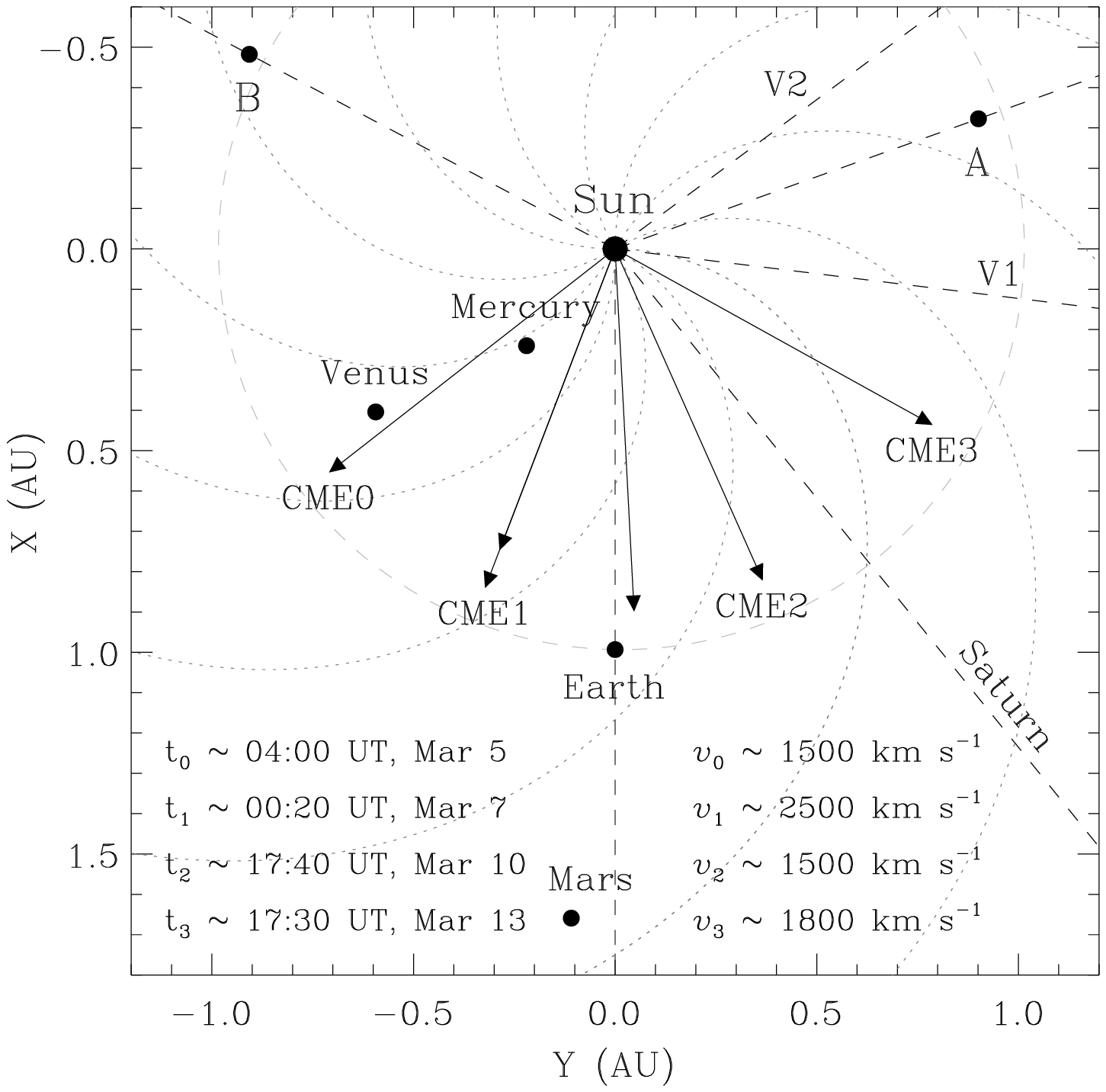} \caption{Positions of the spacecraft and planets in the ecliptic plane on 2012 March 10. The dashed lines indicate the longitudes of the Earth, STEREO A and B, V1 and V2, and Saturn, respectively. The gray dashed circle represents the orbit of the Earth, and the dotted lines show Parker spiral magnetic fields created with a solar wind speed of 450 km s$^{-1}$. The arrows mark the propagation directions of the major CMEs in 2012 March estimated from the longitudes of their source locations on the Sun. The estimated CME speeds and launch times on the Sun are also given.}
\end{figure}

\clearpage

\begin{figure}
\epsscale{0.9} \plotone{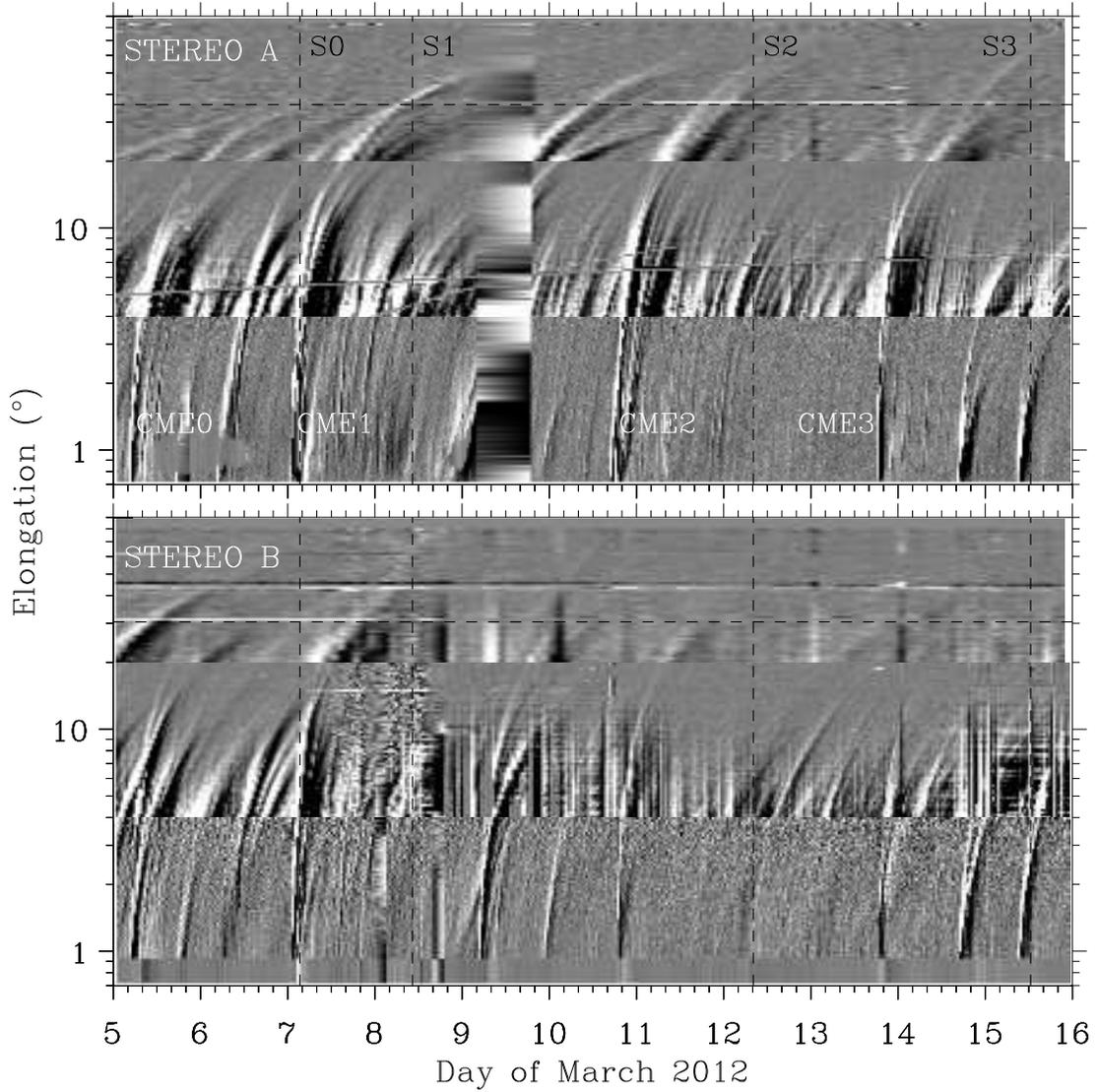} \caption{Time-elongation maps constructed from running-difference images of COR2, HI1 and HI2 along the ecliptic for STEREO A (upper) and B (lower). Tracks associated with the CMEs of interest are indicated. The vertical dashed lines mark the observed shock arrival times at the Earth, and the horizontal dashed line denotes the elongation angle of the Earth. An animation associated with this figure is available in the online journal.}
\end{figure}

\clearpage

\begin{figure}
\epsscale{0.9} \plotone{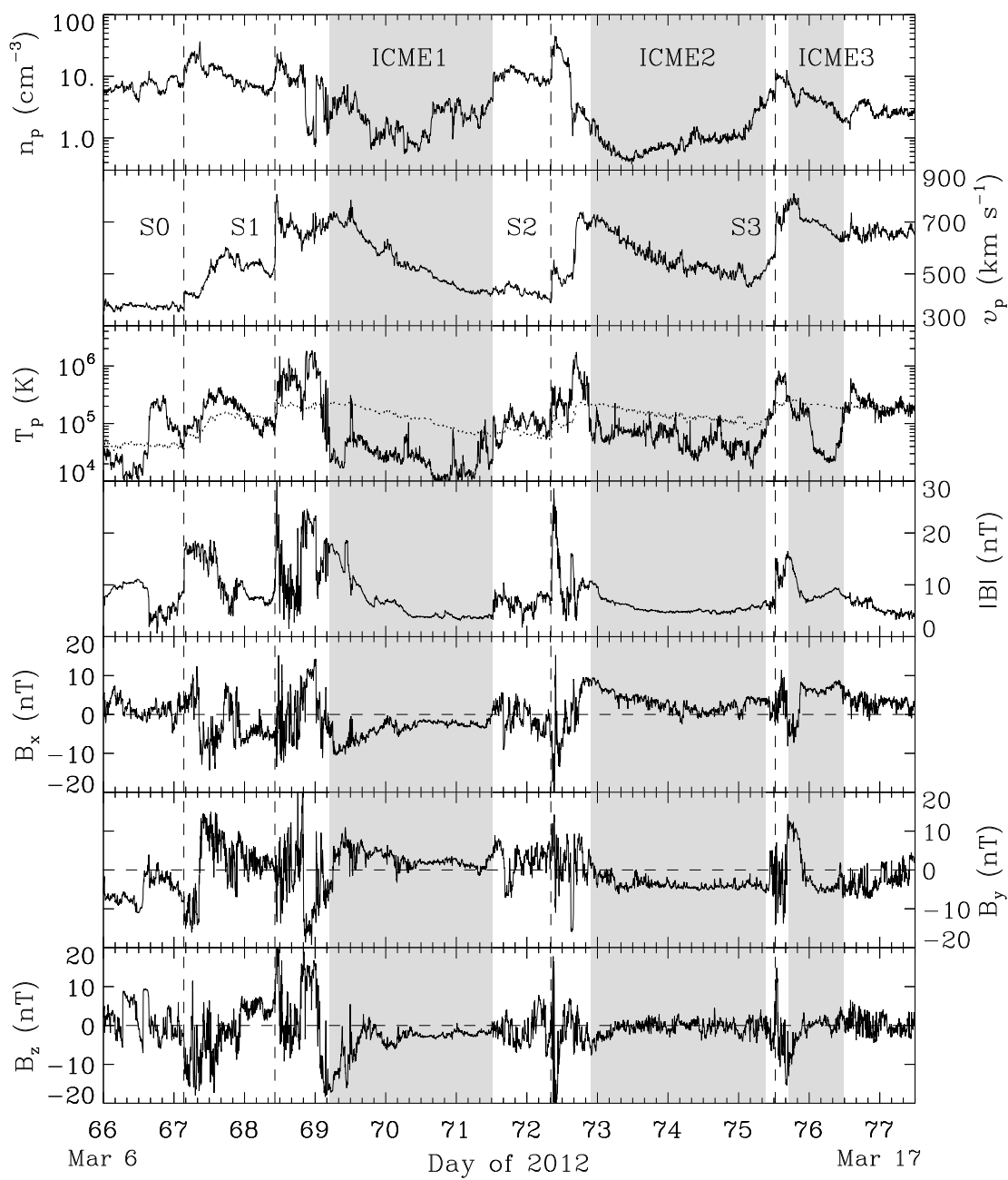} 
\caption{Solar wind plasma and magnetic field parameters observed at Wind. From top to bottom, the panels show the proton density, bulk speed, proton temperature, and magnetic field strength and components, respectively. The dotted curve in the third panel denotes the expected proton temperature from the observed speed. The shaded regions indicate the ICME intervals, and the vertical dashed lines mark the associated shocks.}
\end{figure}

\clearpage

\begin{figure}
\epsscale{0.8} \plotone{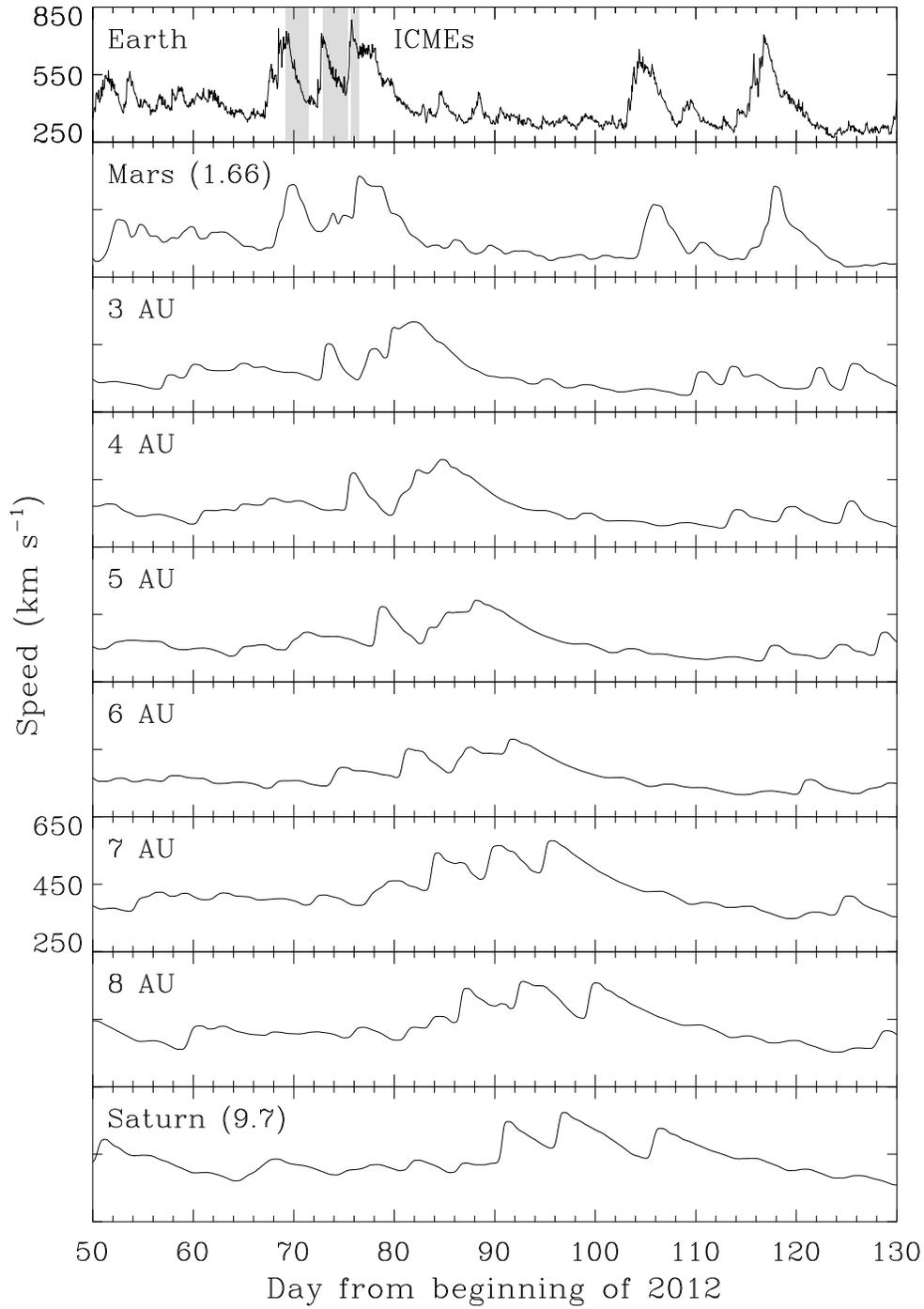} \caption{Evolution of solar wind speeds from the Earth to Saturn (9.7 AU) via the MHD model. The shaded regions represent the ICME intervals at the Earth. The curve in the top panel is the observed speed at Wind, and others are predicted speeds at given distances. Note that the vertical axis is rescaled to [250, 650] km s$^{-1}$ 
at 7 AU and beyond.}
\end{figure}

\clearpage

\begin{figure}
\epsscale{0.8} \plotone{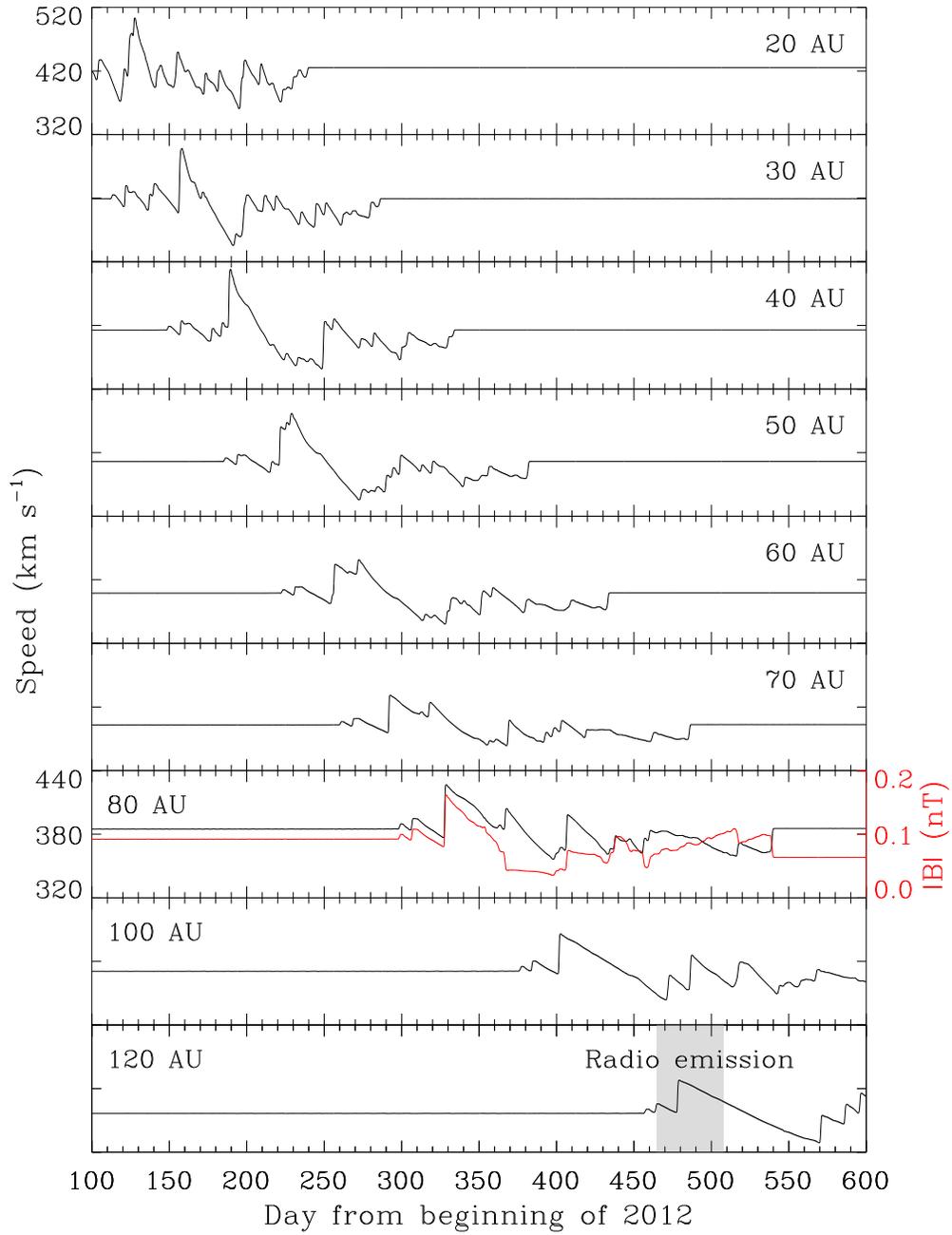} \caption{Similar format to Figure~4, but for the predicted solar wind speeds from 20 to 120 AU. At 80 AU and beyond, the vertical axis is rescaled to [320, 440] km s$^{-1}$. The solar wind magnetic field is also plotted at 80 AU (scaled by the red axis). The shaded region in the bottom panel indicates the period of the radio emission observed by V1.}
\end{figure}

\end{document}